\documentclass[oneside, 11pt]{article}

\usepackage{mystyle}

\newcommand{\E}{\mathbb{E}}
\newcommand{\Var}{\mathrm{Var}}
\newcommand{\Cov}{\mathrm{Cov}}

\newcommand{\Normal}{\mathcal{N}}

\newcommand{\R}{\mathbb{R}}

\DeclarePairedDelimiter{\norm}{\lVert}{\rVert}
\DeclarePairedDelimiter{\abs}{\lvert}{\rvert}

\DeclarePairedDelimiter{\set}{\{}{\}}

\newcommand{\mc}{\mathcal}
\newcommand{\mr}{\mathrm}

\bibliography{example}

\begin{document}

\title{\textbf{Downscaling land surface temperature data using edge detection and block-diagonal Gaussian process regression}\protect\footnotemark[1]}
\author{Sanjit Dandapanthula\protect\footnotemark[2]
    \and
    Margaret Johnson\protect\footnotemark[3]
    \and
    Madeleine Pascolini-Campbell\protect\footnotemark[3] \\[.1em]
    \and 
    Glynn Hulley\protect\footnotemark[3]
    \and
    Mikael Kuusela\protect\footnotemark[2]
    }
\date{\vspace{1em} \today}
\maketitle

\footnotetext[2]{Carnegie Mellon University, Department of Statistics (E-mail: \texttt{\href{mailto:sanjitd@cmu.edu}{sanjitd@cmu.edu}})}
\footnotetext[3]{Jet Propulsion Laboratory, California Institute of Technology}
\footnotetext[1]{This research was partially carried out at the Jet Propulsion Laboratory, California Institute of Technology, under a contract with the National Aeronautics and Space Administration (80NM0018D0004).}

\begin{abstract}
    Accurate and high-resolution estimation of land surface temperature (LST) is crucial in estimating evapotranspiration, a measure of plant water use and a central quantity in agricultural applications. In this work, we develop a novel statistical method for downscaling LST data obtained from NASA's ECOSTRESS mission, using high-resolution data from the Landsat 8 mission as a proxy for modeling agricultural field structure. Using the Landsat data, we identify the boundaries of agricultural fields through edge detection techniques, allowing us to capture the inherent block structure present in the spatial domain. We propose a block-diagonal Gaussian process (BDGP) model that captures the spatial structure of the agricultural fields, leverages independence of LST across fields for computational tractability, and accounts for the change of support present in ECOSTRESS observations. We use the resulting BDGP model to perform Gaussian process regression and obtain high-resolution estimates of LST from ECOSTRESS data, along with uncertainty quantification. Our results demonstrate the practicality of the proposed method in producing reliable high-resolution LST estimates, with potential applications in agriculture, urban planning, and climate studies.
\end{abstract}

\tableofcontents

\section{Introduction}

Obtaining high-resolution estimates of land surface temperature (LST) given observations from remote sensing satellites is an important problem in environmental statistics; for instance, accurate estimation of LST is crucial for understanding the \emph{evapotranspiration} (ET), a central quantity in Earth's water cycle \citep{fisher2021future}. Spatial modeling is also used in this context for gap-filling between observed data, but we do not focus on this application in this work. Current attempts to build statistical models for LST data have not been able to sufficiently account for the complex spatial structure of the data, especially around agricultural fields where estimates of LST are most crucial. Traditional spatial models, such as stationary Gaussian processes, fail to capture the abrupt changes in LST that occur at the boundaries of agricultural fields, which can lead to oversmoothing and inaccurate predictions. Therefore, there is a need for more sophisticated statistical methods that can effectively model the spatial structure present in LST data.

ET is the sum of evaporation and plant transpiration from the Earth's land surface to the atmosphere \citep{fisher2021future}. It is a key component of the water cycle and is a critical variable in the study of climate change, agriculture, and water resource management. ET is influenced by several factors, including LST, solar radiation, wind speed, and humidity. In particular, LST is a key driver of ET as it controls the amount of energy available for evaporation and transpiration. Accurate estimation of LST is therefore essential for predicting ET.

The ECOsystem Spaceborne Thermal Radiometer Experiment on Space Station (ECOSTRESS) radiometer was deployed on the International Space Station (ISS) in June 2018 to measure LST and emissivity, using an infrared sensor \citep{fisher2020ecostress}. The \emph{overpass return frequency}, or revisit rate, of a sensor is the frequency with which that sensor passes over a given location on Earth. The ECOSTRESS mission has a variable overpass return frequency (between 1 and 5 days) and currently provides the highest spatial resolution spaceborne estimates of LST at approximately 70 m resolution. However, the effective resolution of LST images from ECOSTRESS has been shown to be significantly lower in practice \citep{holmes2024orbit}, likely due to blur and artifacts resulting from sensor damage.

The Landsat program is a series of Earth-observing satellite missions jointly managed by NASA and the U.S. Geological Survey (USGS) that has been providing high-resolution imagery of the Earth's surface since 1972 \citep{roy2014landsat}. Landsat satellites are equipped with sensors that capture data in multiple spectral bands, including visible, near-infrared, and thermal infrared, allowing for detailed analysis of land cover, vegetation health, and surface temperature. Landsat 8/9 instruments provide data in the visible to shortwave infrared (VSWIR) at higher resolution (30m) than ECOSTRESS, but with a lower overpass return frequency (16 days). Landsat 8/9 Thermal Infrared Sensors (TIRS) have fewer thermal bands than ECOSTRESS, resulting in potentially lower fidelity estimates of LST, and are at nominally lower spatial resolution (100 m). 

\begin{figure}[htbp]
  \centering
  \begin{subfigure}[b]{0.45\textwidth}
    \centering
    \includegraphics[width=\textwidth,height=4.5cm,keepaspectratio]{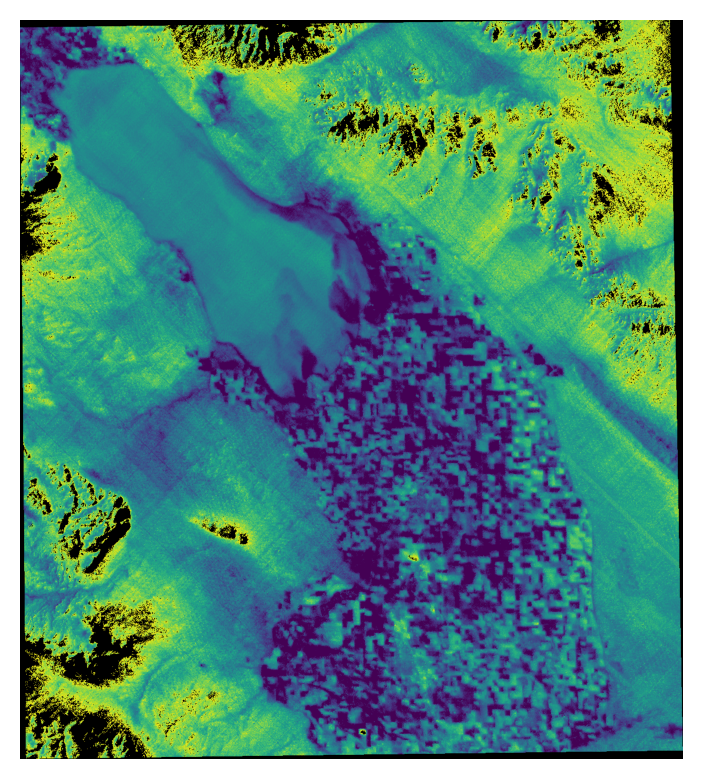}
    \caption{ECOSTRESS (70 m resolution)}
    \label{fig:ecostress_sample}
  \end{subfigure}
  \begin{subfigure}[b]{0.45\textwidth}
    \centering
    \includegraphics[width=\textwidth,height=4.5cm,keepaspectratio]{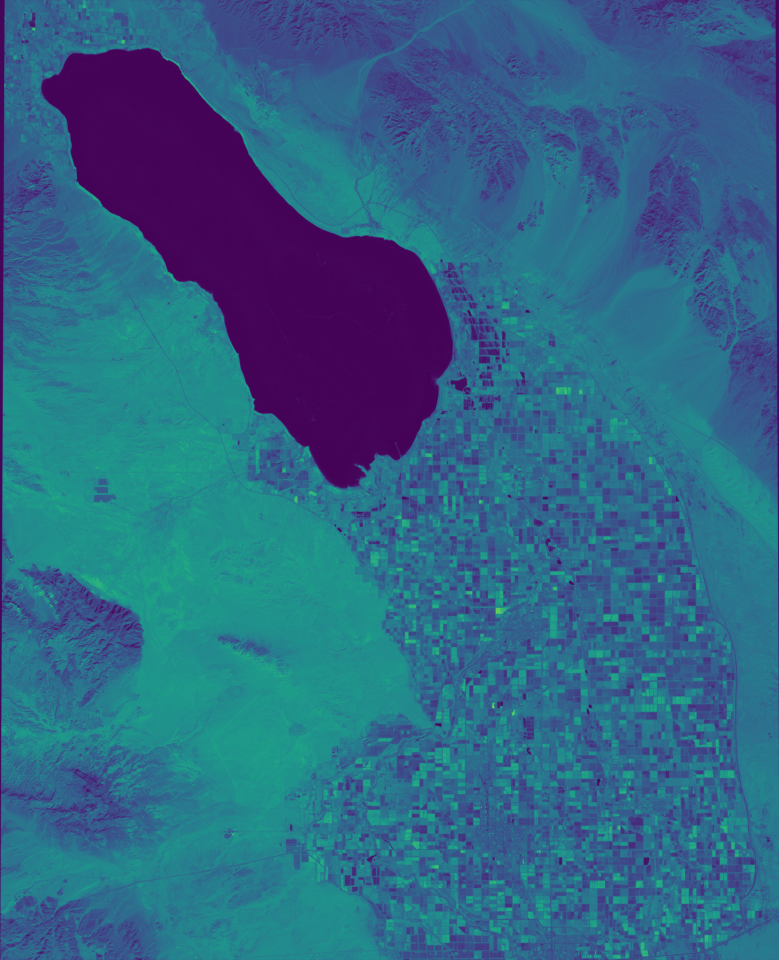}
    \caption{Landsat (30 m resolution)}
    \label{fig:landsat_sample}
  \end{subfigure}
  \caption{Comparison of resolution between ECOSTRESS and Landsat missions over the Salton Sea region in southern California.}
  \label{fig:resolution_comparison}
\end{figure}

In our work, we leverage the higher resolution Landsat 8/9 VSWIR data as a proxy for modeling the agricultural field structure present in the ECOSTRESS LST data. This task, often referred to as \emph{data fusion}, is of broad interest in the remote sensing community, in far more general settings than just LST data \citep{khaleghi2013multisensor, samadzadegan2025critical}. We plot sample images from both missions in \Cref{fig:resolution_comparison}, which reflects the type of agricultural field structure that is the focus of our work. Note that the ECOSTRESS and Landsat images are not in the same units and are not taken at the same time; this figure is only meant to illustrate the difference in spatial resolution.

\vspace{.3cm}

There is a long line of work studying downscaling procedures for land surface temperature data by direct regression of the high-resolution images on their low-resolution counterparts using random forest regression \citep{njuki2020improved, hutengs2016downscaling, li2018downscaling} and neural networks \citep{wu2022downscaling, li2025downscaling, bindhu2013development}. By contrast, our statistical approach using Gaussian process regression is new in this literature. Some key advantages of our approach are the better interpretability of the underlying statistical model, weaker modeling assumptions which may improve robustness and the ability to obtain principled uncertainties for the resulting reconstruction.

More similar to our approach, \citet{garrigues2007using} proposed the mosaic process model for block-structured spatial data, which allows for different covariance structures in different regions of the spatial domain. However, the complexity of their model makes it difficult to scale to larger datasets, such as those encountered in remote sensing applications. Their model is also restricted to blocks that are bounded by straight lines through the spatial domain, which is not ideal for modeling agricultural fields with irregular shapes. Furthermore, the tesselation in the mosaic process model is random which may lead it to be underconstrained for spatial mapping.

Our main contributions in this work are as follows:
\begin{itemize}
  \item We propose a novel statistical method for downscaling LST data from the ECOSTRESS mission using high-resolution data from the Landsat mission as a proxy for modeling agricultural field structure.
  \item We develop a block-diagonal Gaussian process (BDGP) model that captures the spatial structure of the agricultural fields and leverages independence of LST across fields for computational tractability.
  \item We demonstrate the practicality of the proposed method in producing high-resolution LST estimates along with uncertainty quantification.
\end{itemize}

We outline our methodology in \Cref{sec:methods}; the edge detection method is described in \Cref{sec:edge-detection}, the Fourier regression approach for modeling the mean function is described in \Cref{sec:fourier-regression}, the parameter estimation procedure is described in \Cref{sec:parameter-estimation}, and the kriging and uncertainty quantification method is described in \Cref{sec:kriging}. We present our results in \Cref{sec:results} and conclude with a discussion in \Cref{sec:discussion}.


\section{Methods} \label{sec:methods}


It is common to model spatial data using a Gaussian process (GP) \citep{gelfand2010handbook}. In order to account for the agricultural field structure present in LST data, we model LST using a block-diagonal Gaussian process with a squared exponential kernel. Suppose we have a partition, $\mc{R} \coloneq \set{R_1, \dots, R_N}$, of $\mc{X}$ into $N$ regions. Given a variance parameter $\bm{\sigma} \coloneq (\sigma_1, \dots, \sigma_N) \in (0, \infty)^N$ and length scale parameter $\bm{\ell} \coloneq (\ell_1, \dots, \ell_N) \in (0, \infty)^N$, we define the block-diagonal squared exponential covariance kernel with parameters $(\bm{\sigma}, \bm{\ell})$ as $k(x, x^\prime) = \sum_{r=1}^N k_r(x,x^\prime)$ where:

\begin{align*}
  k_r(x, x^\prime) 
  = \begin{cases}
      \sigma_r^2\, \exp\left( -\norm{x - x^\prime}_2^2 / (2 \ell_r^2) \right), & \quad x, x^\prime \in R_r \\
      0,                                                                       & \quad \text{otherwise}.
    \end{cases}
\end{align*}
This kernel is stationary within each region. Given any mean function $\mu : \mc{X} \to \R$, we say that a block diagonal Gaussian process with parameters $(\mu, \bm{\sigma}, \bm{\ell}, \mc{R})$ (denoted $\mathrm{BDGP}(\mu, \bm{\sigma}, \bm{\ell}, \mc{R})$) is a GP with mean function $\mu$ and covariance kernel $k$ of the form above.

We assume that at a given time the underlying LST field is a BDGP $f^\mathrm{ES} \coloneq \mathrm{BDGP}(\mu^\mathrm{ES}, \sigma^\mathrm{ES}, \ell^\mathrm{ES}, \mc{R})$ and that we observe samples from the blurry $\tilde{f}^\mathrm{ES} \coloneq (f^\mathrm{ES} \ast \kappa_\mathrm{blur}) + f_\mathrm{sensor}$, where $\ast$ denotes a two-dimensional convolution. Here, $\kappa_\mathrm{blur}$ is the density of $\Normal(0,\, \sigma_\mathrm{blur}^2\, I)$ (accounting for the change of support caused by the finite resolution of ECOSTRESS) and $f_\mathrm{sensor} \sim \Normal(0,\, \sigma_\mathrm{sensor}^2\, I)$ accounts for sensor noise within ECOSTRESS. We further assume that the Landsat data are generated from $f^\mathrm{LS} \coloneq \mathrm{BDGP}(\mu^\mathrm{LS}, \sigma^\mathrm{LS}, \ell^\mathrm{LS}, \mc{R})$. We make the simplifying assumption that $\ell^\mathrm{ES} = \ell^\mathrm{LS}$ in order to ensure that the parameters are identifiable in this model.

Then, we outline our LST downscaling procedure as follows:
\begin{itemize}
  \item[(1)] \emph{Edge detection}: Use edge detection methods on an averaged high-resolution Landsat image to identify boundaries between different agricultural fields, thereby estimating the partition $\mc{R}$.
  \item[(2)] \emph{Parameter estimation}: Fixing $\mc{R}$, fit the remaining parameters of the BDGP model to the Landsat and ECOSTRESS data by maximum likelihood estimation.
  \item[(3)] \emph{Kriging and uncertainty quantification}: Use the fitted BDGP model for ECOSTRESS to do kriging and obtain high-resolution predictions of the residual LST field ($f^\mr{ES} - \E[f^\mr{ES}]$), along with uncertainty quantification.
\end{itemize}

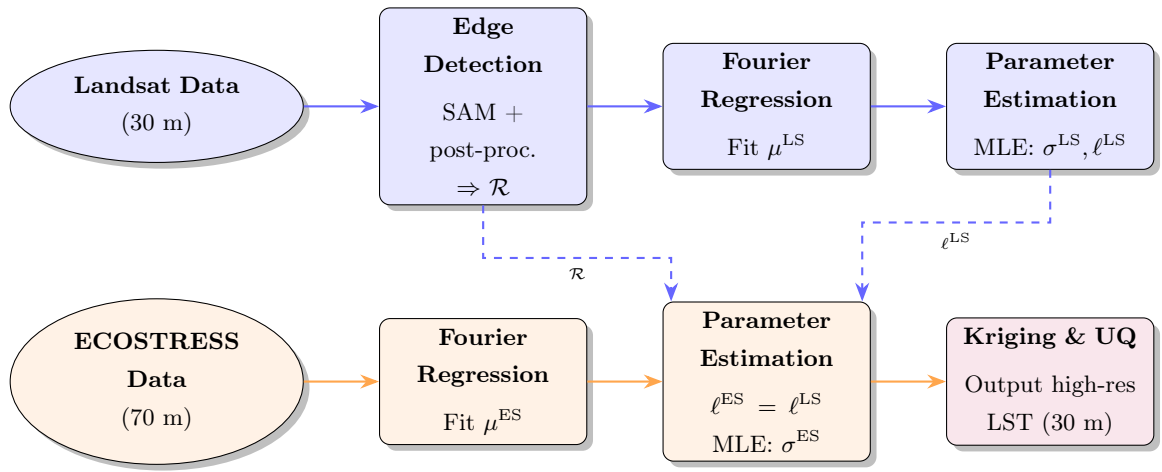
\begin{figure}[htbp]
  \centering
  \begin{tikzpicture}[
    node distance=1cm and 1cm,
    block/.style={rectangle, draw, fill=blue!15, text width=2.5cm, text centered, rounded corners, minimum height=1.3cm, drop shadow, font=\footnotesize},
    data/.style={ellipse, draw, fill=green!15, text width=2.5cm, text centered, minimum height=0.9cm, drop shadow, font=\footnotesize},
    arrow/.style={-{Stealth[length=2.5mm]}, thick, draw=gray!70},
    arrowblue/.style={-{Stealth[length=2.5mm]}, thick, draw=blue!60},
    arroworange/.style={-{Stealth[length=2.5mm]}, thick, draw=orange!70}
    ]

    \node[data, fill=blue!10] (landsat) {\textbf{Landsat Data}\\(30 m)};
    \node[data, fill=orange!10, below=1.8cm of landsat] (ecostress) {\textbf{ECOSTRESS Data}\\(70 m)};

    \node[block, right=of landsat, fill=blue!10] (edge) {\textbf{Edge \\Detection}\\[1ex]SAM + post-proc.\\$\Rightarrow$ $\mathcal{R}$};
    \node[block, right=of edge, fill=blue!10] (fourier-ls) {\textbf{Fourier\\ Regression}\\[1ex]Fit $\mu^{\mathrm{LS}}$};
    \node[block, right=of fourier-ls, fill=blue!10] (params-ls) {\textbf{Parameter Estimation}\\[1ex]MLE: $\sigma^{\mathrm{LS}}, \ell^{\mathrm{LS}}$};

    \node[block, right=of ecostress, fill=orange!10] (fourier-es) {\textbf{Fourier\\ Regression}\\[1ex]Fit $\mu^{\mathrm{ES}}$};
    \node[block, right=of fourier-es, fill=orange!10] (params-es) {\textbf{Parameter Estimation}\\[1ex]$\ell^{\mathrm{ES}} = \ell^{\mathrm{LS}}$\\MLE: $\sigma^{\mathrm{ES}}$};

    \node[block, right=of params-es, fill=purple!10] (kriging) {\textbf{Kriging \& UQ}\\[1ex] Output high-res LST (30 m)};


    \draw[arrowblue] (landsat) -- (edge);
    \draw[arrowblue] (edge) -- (fourier-ls);
    \draw[arrowblue] (fourier-ls) -- (params-ls);

    \draw[arroworange] (ecostress) -- (fourier-es);
    \draw[arroworange] (fourier-es) -- (params-es);

    \draw[arrowblue, dashed] (edge.south) -- ++(0,-0.7) -| node[pos=0.25, below, font=\tiny] {$\mathcal{R}$} (params-es.140);
    \draw[arrowblue, dashed] (params-ls.south) -- ++(0,-0.7) -| node[pos=0.25, below, font=\tiny] {$\ell^{\mathrm{LS}}$} (params-es.40);

    \draw[arroworange] (params-es) -- (kriging);
  \end{tikzpicture}
  \caption{Overview of the proposed downscaling pipeline. Overview of the proposed downscaling pipeline. The Landsat data is shown in blue, the ECOSTRESS data is shown in orange, and the output of the algorithm  is shown in purple.}
  \label{fig:pipeline}
\end{figure}

An overview of our downscaling pipeline is depicted in \Cref{fig:pipeline}.

\vspace{.3cm}

\subsection{Edge detection} \label{sec:edge-detection}

First, we average 6 separate high-resolution Landsat images and use edge detection methods on the averaged image to identify boundaries between different agricultural fields. In our case, we use Meta's Segment Anything Model (SAM), which is a state-of-the-art vision transformer (ViT) based segmentation model providing an over-segmentation of the input image---the model may produce partially or fully overlapping segments \citep{kirillov2023segment}. We experimented with several other methods, including the Canny, Sobel, and Laplace edge detection algorithms, but we found the SAM to work best in practice. Next, we post-process the over-segmented image to merge overlapping regions together using an algorithm described in \Cref{app:combination-algorithm}. This post-processing step ensures that the final set of segments contains no overlaps. At this point, we have an estimated partition $\mc{R}$ of the spatial domain into agricultural fields.

\subsection{Fourier regression} \label{sec:fourier-regression}

Next, we fit $\mu^\mathrm{LS}$ separately for each pixel $x = (i, j)$ using a Fourier regression with annual frequency:
\begin{align*}
  \mu_{ij}^\mathrm{LS}(t)
  = (\beta_0)_{ij} + (\beta_1)_{ij} \cos(2 \pi t / 365) + (\beta_2)_{ij} \sin(2 \pi t / 365),
\end{align*}
where $t$ is in units of days. We use ordinary least squares to fit the coefficients. We only use an annual frequency here, because the Landsat satellite has a \emph{sun-synchronous} orbit, meaning that the satellite orbits over the same spot on Earth at the same time of day every 16 days. As a result, we are unable to model the diurnal frequency for these data. Next, we fit the blurred mean $\mu^\mathrm{ES} \ast \kappa_\mathrm{blur}$ separately for each pixel $(i, j)$ using a Fourier regression with annual and diurnal frequencies:
\begin{align*}
   & (\mu^\mathrm{ES} \ast \kappa_\mathrm{blur})_{ij}(t)                                                                                                            \\
   & \quad = (\beta_0)_{ij} + (\beta_1)_{ij} \cos(2 \pi t / 365) + (\beta_2)_{ij} \sin(2 \pi t / 365) + (\beta_3)_{ij} \cos(2 \pi t) + (\beta_4)_{ij} \sin(2 \pi t)
\end{align*}
where $t$ is in units of days. Note that we are fitting the blurred mean function here since we only observe the blurred data from ECOSTRESS. At last, we subtract the fitted mean functions from the observed data, so that we may model the residual data by zero-mean GPs. We see in \Cref{fig:fourier_fit_pixel} (\Cref{app:additional-figures}) that a single Fourier term is sufficient to capture the annual and diurnal cycles in the data, so that we can model the residuals by a mean-zero BDGP. Similarly, we depict the annual cycle fit for a random pixel in the Landsat data in \Cref{fig:annual_cycle_landsat} (\Cref{app:additional-figures}). The fitted cycles roughly capture the seasonal variations in land surface temperature, justifying our use of Fourier regression to model the mean function.

\subsection{Parameter estimation} \label{sec:parameter-estimation}

Since the uncertainties listed for the ECOSTRESS observations are all on the order of 0.1 degrees Kelvin, it is reasonable to choose $\sigma_\mathrm{sensor} = 0.1$. From \citep{holmes2024orbit}, we know that the point spread function of ECOSTRESS approximately has a full-width at half-maximum (FWHM) of 160 m in each direction, which is roughly twice the nominal pixel size of 70 m. From the FWHM, we can calculate the standard deviation by $\sigma_\mathrm{blur} = (160 / 70) / (2 \sqrt{2 \ln(2)}) \approx 0.97$ in units of ECOSTRESS pixels. This computation is justified in \citet{fwhm}.

Fixing $\mc{R}$, we now fit the remaining parameters of the BDGP model to the Landsat data. Note that each region can be fit independently here, which allows for easy parallelization and scalability to larger data. We do not include a temporal component in the covariance structure here, and instead sample far apart in time to reduce temporal correlation. As a result, we can treat these points as independent across time. We choose 4 sample dates $t_1, t_2, t_3, t_4$ that are approximately three months apart from each other. We write the likelihood of the observed data in region $R_r$ assuming that the observations are independent across time:
\begin{align*}
  \mc{L}(\sigma_r^\mathrm{LS}, \ell_r^\mathrm{LS}; y)
  = \prod_{i=1}^4 f_{\Normal(0, K_{R_r})}(y_{R_r, t_i}^\mathrm{LS}),
\end{align*}
where $K_{R_r}$ is the covariance matrix of the BDGP restricted to region $r$ and $y_{R_r, t_i}^\mathrm{LS}$ are the observed residuals in region $R_r$ at time $t_i$. We estimate the parameters by maximizing this likelihood using the BFGS optimizer, adding a small nugget of $10^{-8} I$ to the covariance matrix for numerical stability.

Next, we fix $\mc{R}$ and assume that $\ell^\mathrm{ES} = \ell^\mathrm{LS}$. Although the Landsat and ECOSTRESS data measure different quantities, it is reasonable to think that the length scale of the residual field should be similar across both datasets since they are over the same geographical region with similar spatial variation. This assumption is required to ensure identifiability of the parameters in our model, due to ill-posedness caused by the change of support present in the ECOSTRESS data. Then, we fit $\sigma^\mathrm{ES}$ to the ECOSTRESS data by maximum likelihood estimation.

We now ignore the existence of blurring from other regions at the edges of each region, and fit each region independently; this is justified due to the small relative scale of the blur kernel compared to the regions. Under this simplifying assumption, a consistent estimator given a single observation is
\begin{equation} \label{eq:variance-estimator}
  (\hat{\sigma}^\mathrm{ES}_r)^2
  = (\widehat{\Var}(y_{R_r}^\mathrm{ES}) - \sigma_\mathrm{sensor}^2) \left( \frac{(\ell_r^\mathrm{ES})^2 + \sigma_\mathrm{blur}^2}{(\ell_r^\mathrm{ES})^2} \right).
\end{equation}
Here, we use $\Var(y_{R_r}^\mathrm{ES})$ to denote the sample variance of the observed residuals in region $R_r$, and we justify our use of this estimator in \Cref{app:variance-est}. We then aggregate our $\sigma_r^\mathrm{ES}$ estimates from each separate time point by taking the median across times.

\subsection{Kriging and uncertainty quantification} \label{sec:kriging}

Finally, we use the fitted BDGP model for ECOSTRESS to do kriging (on a 30 m grid) and obtain high-resolution predictions of the residual LST field. Although the regions are no longer independent after blurring, we may treat each region independently after considering a $4 \sigma_\mathrm{blur}$-pixel neighborhood $R_r^\prime$ around each region $R_r$. First, enumerate the pixels as $x_1, \dots, x_P$ and define $k_\ast$ as the covariance vector between $f(x_\ast)$ and all blurry observations in $R_r^\prime$:
\begin{align*}
  (k_\ast)_i = \Cov(f^\mathrm{ES}(x_\ast),\, \tilde{f}^\mathrm{ES}(x_i)) = (k(x_\ast,\, \cdot) \ast \kappa_\mathrm{blur}(\cdot))(x_i).
\end{align*}
Next, let $K_{R_r^\prime}$ be the covariance matrix of the BDGP restricted to region $R_r^\prime$:
\begin{align*}
  (K_{R_r^\prime})_{ij} = \Cov(\tilde{f}^\mathrm{ES}(x_i),\, \tilde{f}^\mathrm{ES}(x_j)) = \iint \kappa_\mathrm{blur}(x_i - u)\, \kappa_\mathrm{blur}(x_j - v)\, k(u, v)\, du\, dv.
\end{align*}
Finally, let $y_{R_r^\prime}^\mathrm{ES}$ be the observed residuals in region $R_r^\prime$; namely, $(y_{R_r^\prime}^\mathrm{ES})_i = \tilde{f}^\mathrm{ES}(x_i)$ for $x_i \in R_r^\prime$. Using this notation, the kriging formula \citep{cressie2015statistics} gives us the conditional mean in region $R_r$ at location $x_\ast$ as
\begin{align*}
  \hat{f}^\mathrm{ES}_{R_r}(x_\ast)
  = k_\ast^\top (K_{R_r^\prime} + \sigma_\mathrm{sensor}^2 I)^{-1} y_{R_r^\prime}^\mathrm{ES}.
\end{align*}
Repeating this computation for each region $R_r$ gives us the full kriging prediction $\hat{f}^\mathrm{ES}$. By reconstructing each region separately, we can parallelize this step and maintain computational tractability as long as each region is not too large. In this work, we do not reconstruct the road region, since it is too large to reconstruct efficiently using our current implementation. Although the roads can be reconstructed by splitting into several overlapping sub-regions, the roads are not of primary interest in agricultural applications, so we leave this for future work. Additionally, because we still estimate the parameters for the road region, this does not affect the reconstructions in other regions.

Next, to compute the variance of the kriging prediction at location $x_\ast$ in region $R_r$, we use the formula
\begin{align*}
  \Var(\hat{f}^\mathrm{ES}(x_\ast))
  = \Var(f^\mathrm{ES}(x_\ast)) - k_\ast^\top (K_{R_r^\prime} + \sigma_\mathrm{sensor}^2\, I)^{-1} k_\ast.
\end{align*}
For instance, this formula can be found in \citet{cressie2015statistics}. Note that the resulting reconstruction is only of the \emph{residual} field; to obtain the full LST field, we would need to add back a deconvolved mean function, which can be done more carefully using methods from spatial statistics. However, we leave this for future work.

\section{Results} \label{sec:results}

For our analysis, we consider an agricultural region of 6 $\mathrm{km}^2$ near the Salton Sea in southern California (USA). We use data from the SR\_B5 channel of the Landsat 8 satellite \citep{roy2014landsat}, which measures surface reflectance in the near-infrared spectrum of light. Our goal is to deconvolve an ECOSTRESS \citep{fisher2020ecostress} LST image taken in January 2021 using the Landsat data as a proxy for modeling the agricultural field structure. We preprocess both datasets to be on a common grid with 30 m resolution. For all of our analysis, we manually choose images which are relatively cloud-free in our region of interest, and do not account for missingness. All of our parameter estimation and kriging procedures work with slight modifications to the code in the presence of some missing data. Similarly, large amounts of missingness can be handled by including a temporal component in the Gaussian process, but this is left for future work.

\begin{figure}[htbp]
  \centering
  \begin{subfigure}[b]{0.32\textwidth}
    \centering
    \includegraphics[width=\textwidth,height=4.5cm,keepaspectratio]{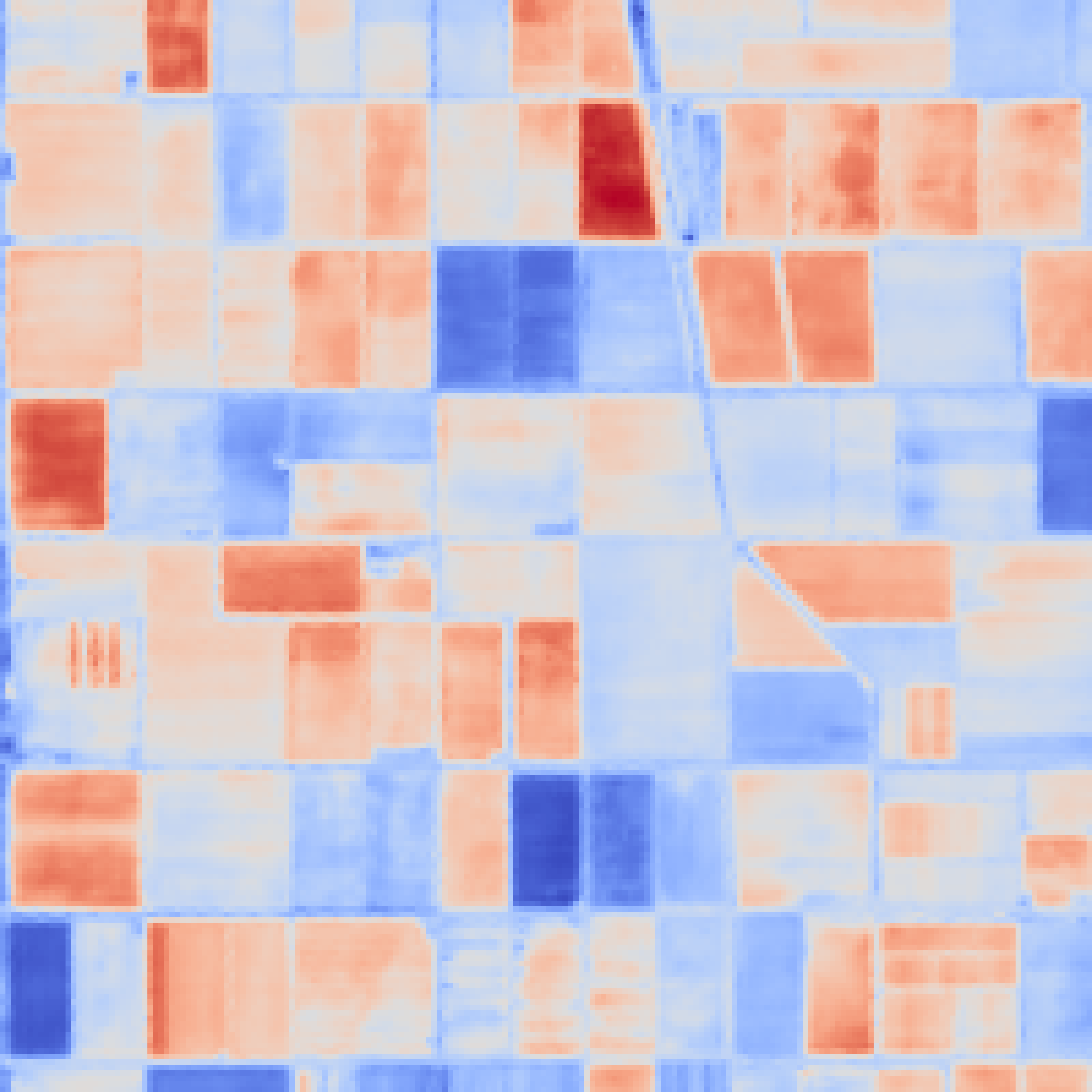}
    \caption{Averaged Landsat image}
    \label{fig:landsat_original}
  \end{subfigure}
  \begin{subfigure}[b]{0.32\textwidth}
    \centering
    \includegraphics[width=\textwidth,height=4.5cm,keepaspectratio]{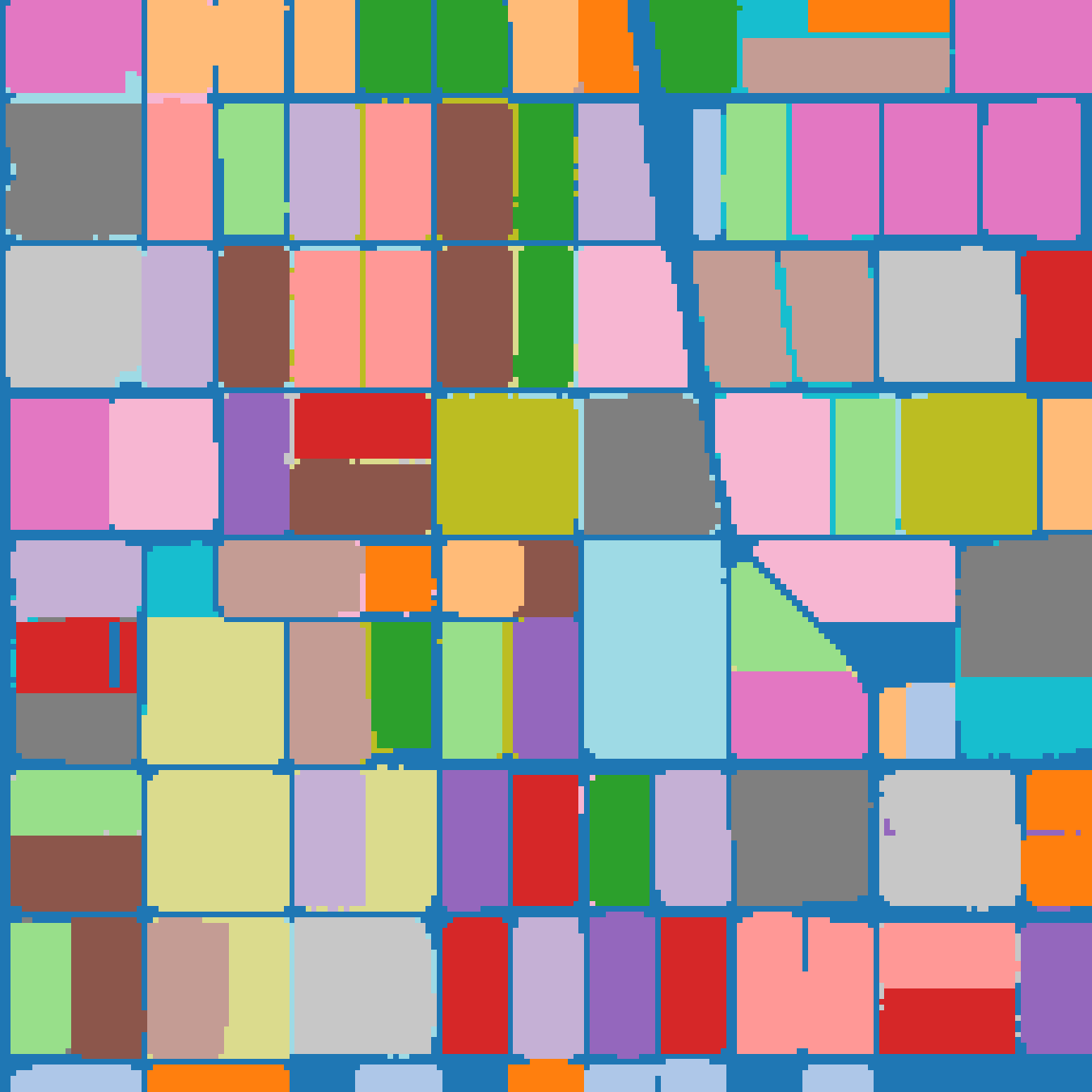}
    \caption{SAM oversegmentation}
    \label{fig:sam_oversegmented}
  \end{subfigure}
  \begin{subfigure}[b]{0.32\textwidth}
    \centering
    \includegraphics[width=\textwidth,height=4.5cm,keepaspectratio]{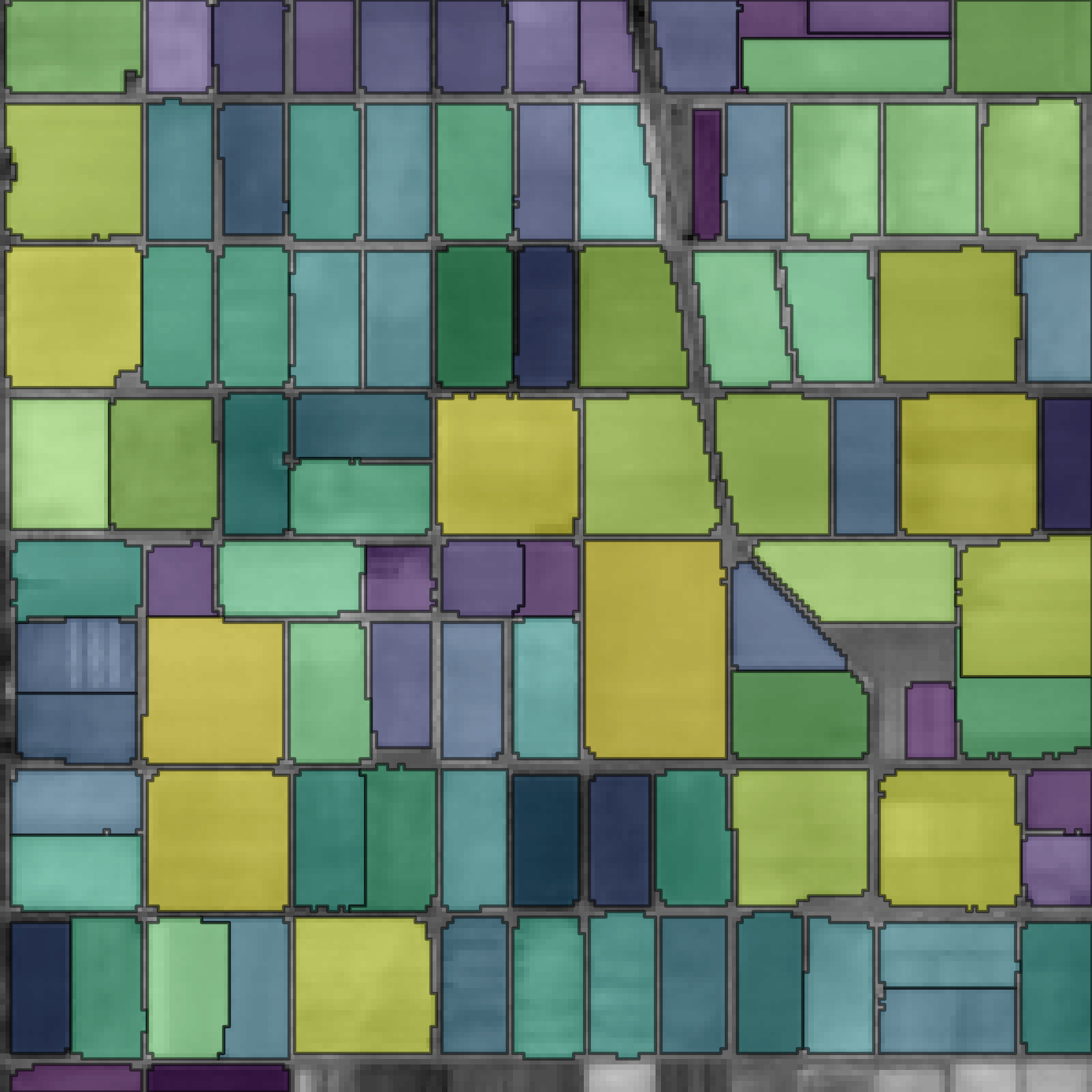}
    \caption{Regions after post-processing}
    \label{fig:final_regions}
  \end{subfigure}
  \caption{Side-by-side comparison of (a) the averaged Landsat image, (b) the oversegmented image from SAM, and (c) the final regions after post-processing.}
  \label{fig:edge_detection_results}
\end{figure}

We begin by showing the results of the edge detection method on the average of 6 Landsat images (starting in November 2021 and spread out by approximately two months) in our region of interest in \Cref{fig:edge_detection_results}. Hence, we confirm that the final regions capture the agricultural field structure relatively well, and there are no overlaps between different regions after post-processing.

\vspace{.3cm}

\begin{figure}[htbp]
  \centering
  \begin{subfigure}[b]{0.24\textwidth}
    \centering
    \includegraphics[width=\textwidth,height=3.5cm,keepaspectratio]{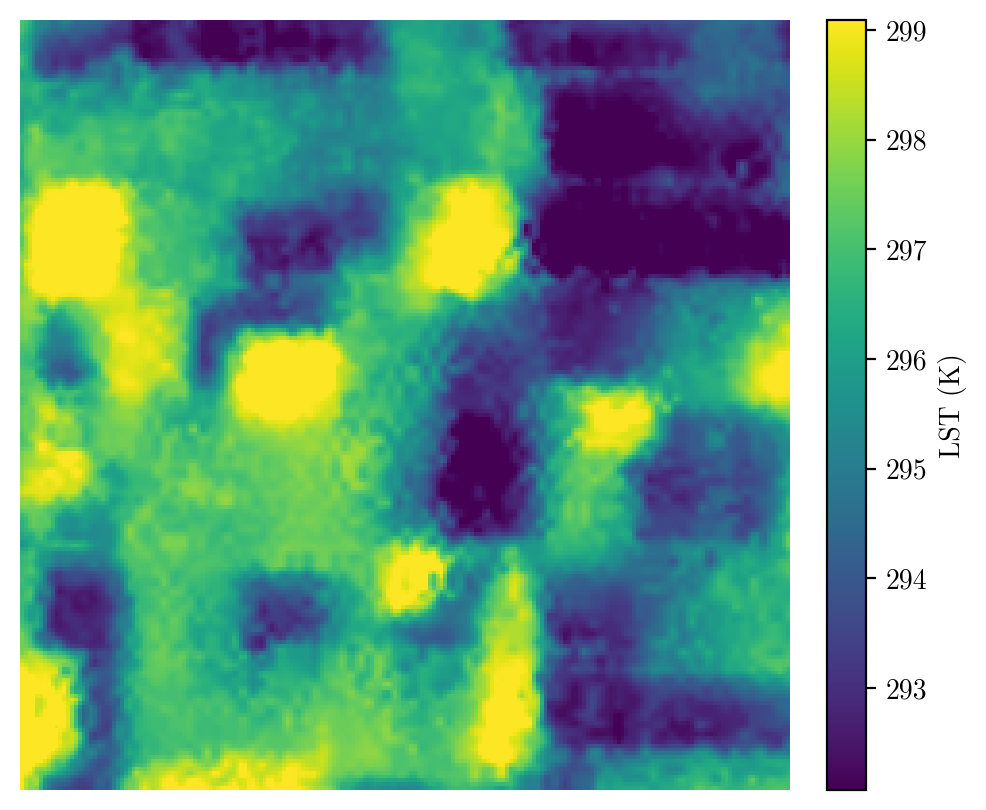}
    \caption{ECOSTRESS original}
    \label{fig:ecostress_original}
  \end{subfigure}
  \begin{subfigure}[b]{0.24\textwidth}
    \centering
    \includegraphics[width=\textwidth,height=3.5cm,keepaspectratio]{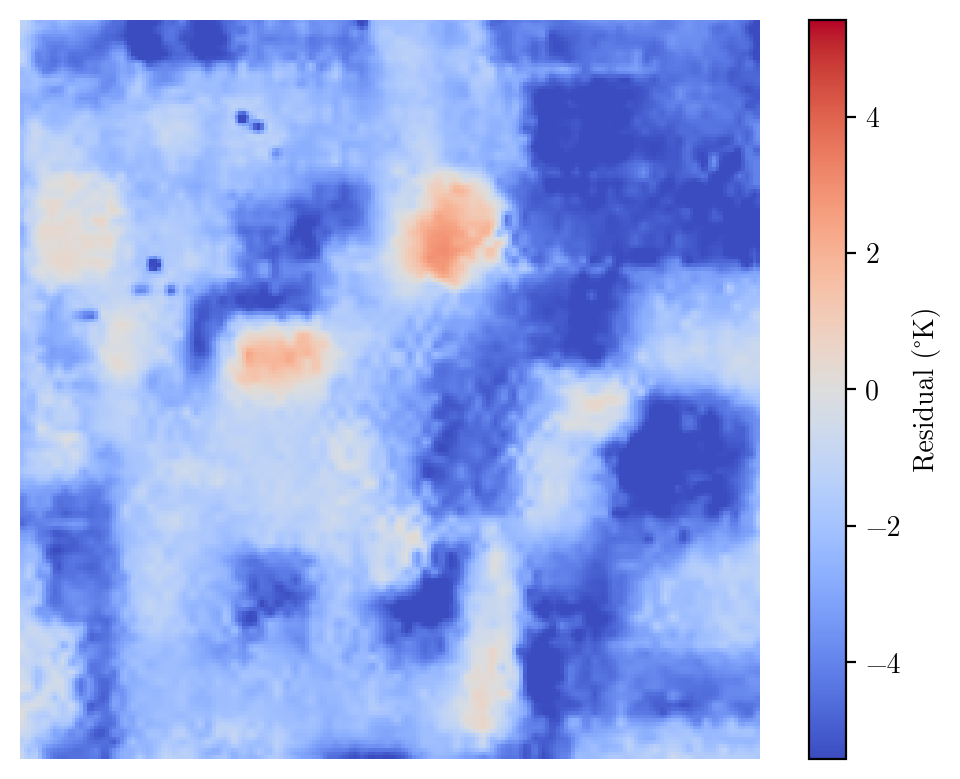}
    \caption{ECOSTRESS residuals}
    \label{fig:ecostress_residuals}
  \end{subfigure}
  \begin{subfigure}[b]{0.24\textwidth}
    \centering
    \includegraphics[width=\textwidth,height=3.5cm,keepaspectratio]{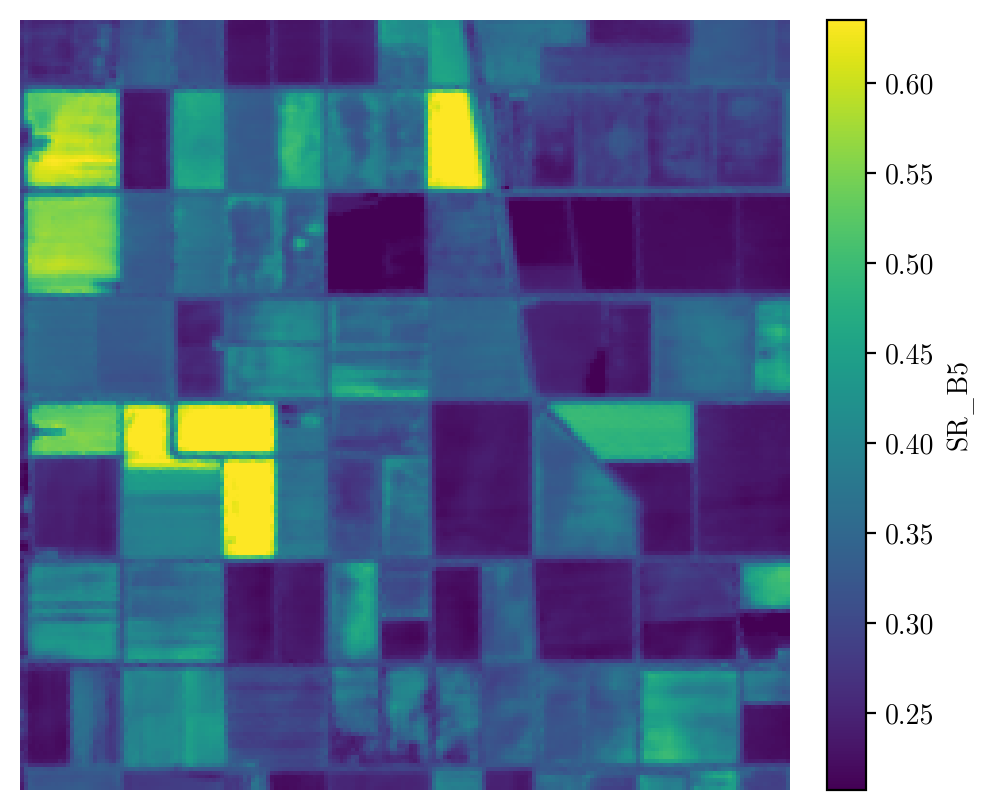}
    \caption{Landsat original}
    \label{fig:landsat_original_sample}
  \end{subfigure}
  \begin{subfigure}[b]{0.24\textwidth}
    \centering
    \includegraphics[width=\textwidth,height=3.5cm,keepaspectratio]{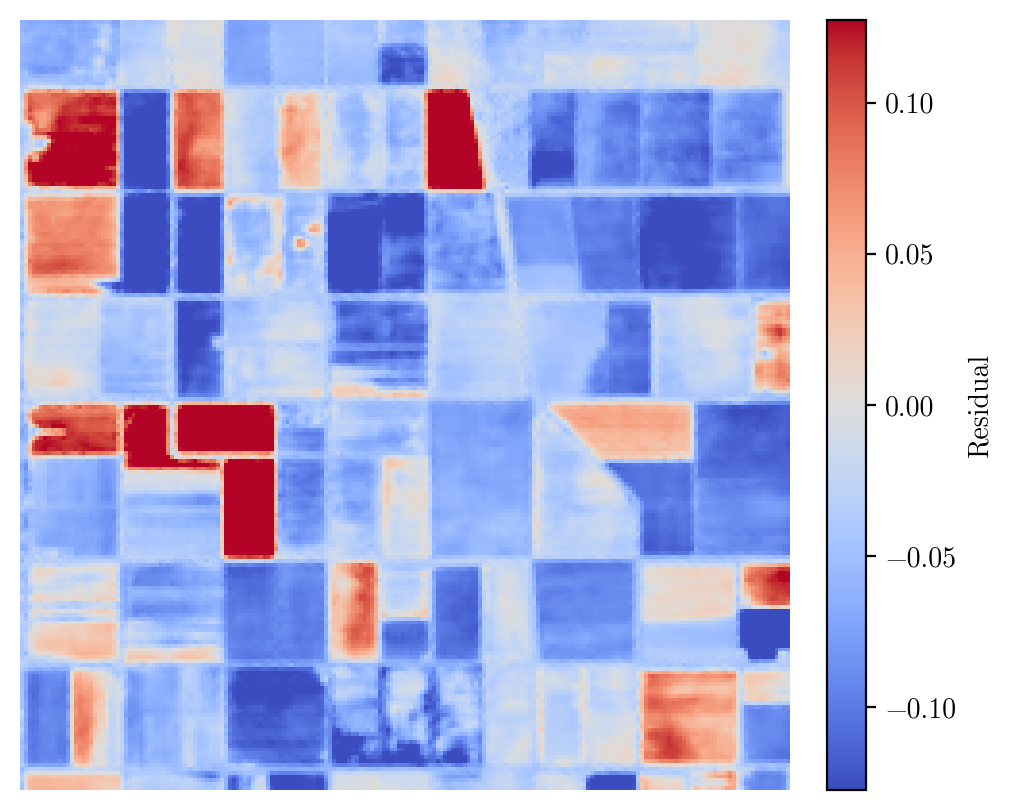}
    \caption{Landsat residuals}
    \label{fig:landsat_residuals}
  \end{subfigure}
  \caption{Side-by-side comparison of (a) the original ECOSTRESS image, (b) the ECOSTRESS residuals after subtracting the fitted mean function, (c) the original Landsat image, and (d) the Landsat residuals after subtracting the fitted mean function.}
  \label{fig:ecostress_landsat_comparison}
\end{figure}

Next, we show our original ECOSTRESS image from January 2021 in \Cref{fig:ecostress_original} (which we would like to deblur) along with the resulting residuals after subtracting the fitted mean function in \Cref{fig:ecostress_residuals}. As expected, the residuals contain significant spatial structure that we can model using our BDGP approach. This day was particularly cold, leading to negative residuals across most of the image. To verify that our mean function is sensible for the Landsat data, we also show the residuals for a sample Landsat image in \Cref{fig:landsat_residuals}. Again, we see that the residuals contain significant spatial structure that we can model using our BDGP approach, and are close to zero.

\vspace{.3cm}

\begin{figure}[htbp]
  \centering
  \begin{subfigure}[b]{0.32\textwidth}
    \centering
    \includegraphics[width=\textwidth,height=4.5cm,keepaspectratio]{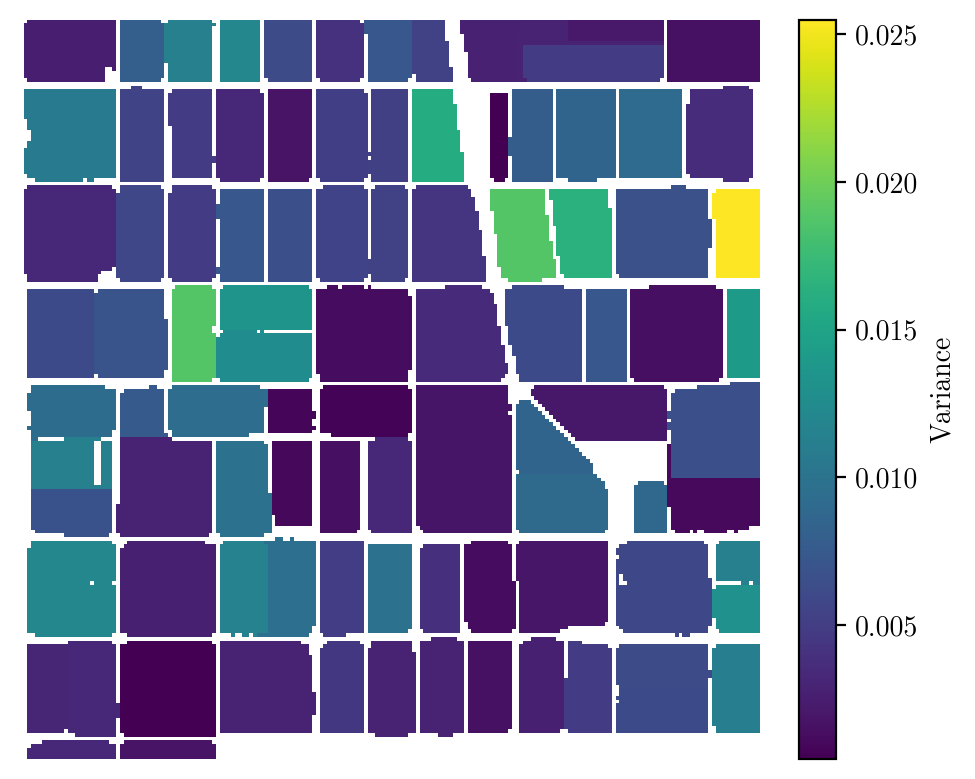}
    \caption{Landsat per-region variance}
    \label{fig:landsat_variance}
  \end{subfigure}
  \begin{subfigure}[b]{0.32\textwidth}
    \centering
    \includegraphics[width=\textwidth,height=4.5cm,keepaspectratio]{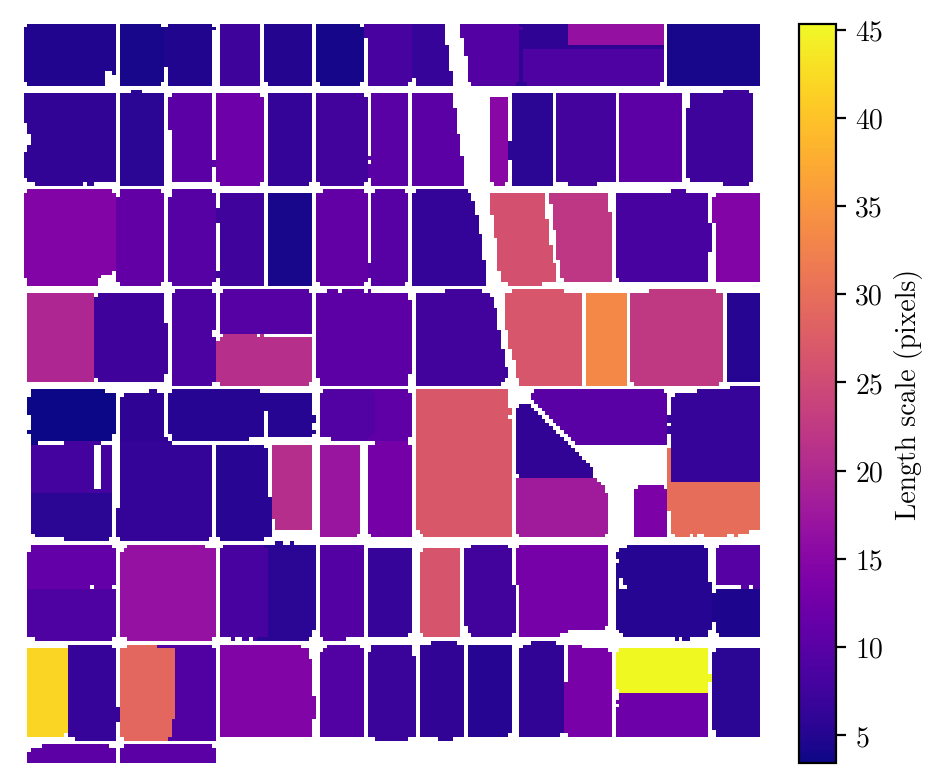}
    \caption{Per-region length scales}
    \label{fig:length_scales}
  \end{subfigure}
  \begin{subfigure}[b]{0.32\textwidth}
    \centering
    \includegraphics[width=\textwidth,height=4.5cm,keepaspectratio]{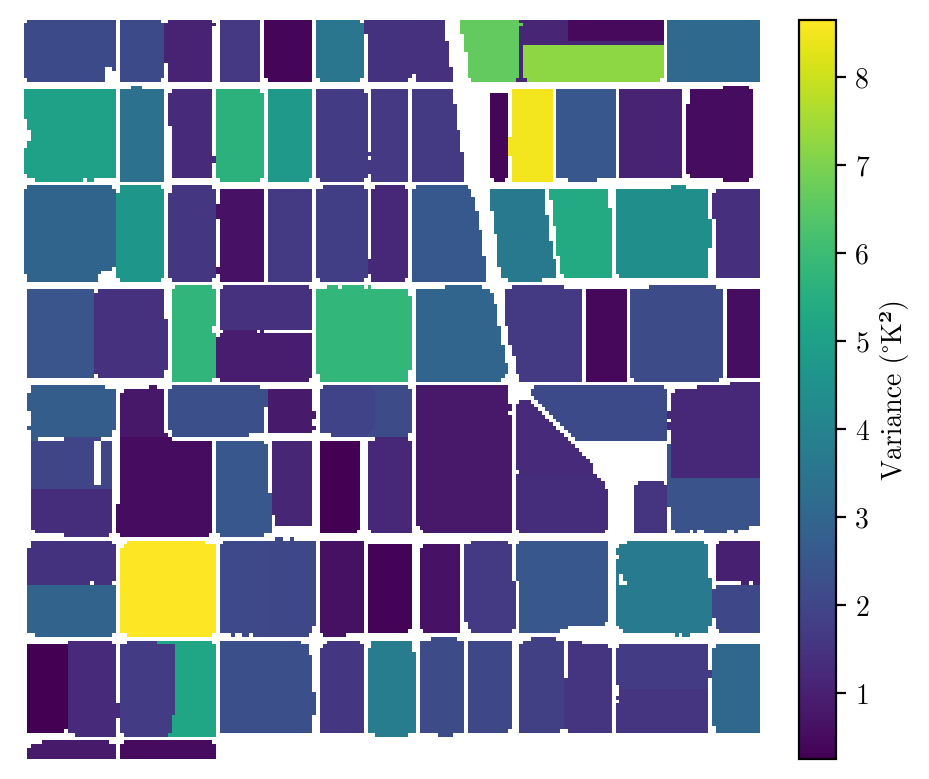}
    \caption{ECOSTRESS per-region variance}
    \label{fig:ecostress_variance}
  \end{subfigure}
  \caption{Fitted parameters for each region: (a) Landsat per-region variance, (b) per-region length scales (shared between Landsat and ECOSTRESS), and (c) ECOSTRESS per-region variance.}
  \label{fig:learned_parameters}
\end{figure}
\begin{figure}[htbp]
  \centering
  \begin{subfigure}[b]{0.32\textwidth}
    \centering
    \includegraphics[width=\textwidth,height=4.5cm,keepaspectratio]{figures/ecostress-residual}
    \caption{Original ECOSTRESS residual}
    \label{fig:ecostress_residual}
  \end{subfigure}
  \begin{subfigure}[b]{0.32\textwidth}
    \centering
    \includegraphics[width=\textwidth,height=4.5cm,keepaspectratio]{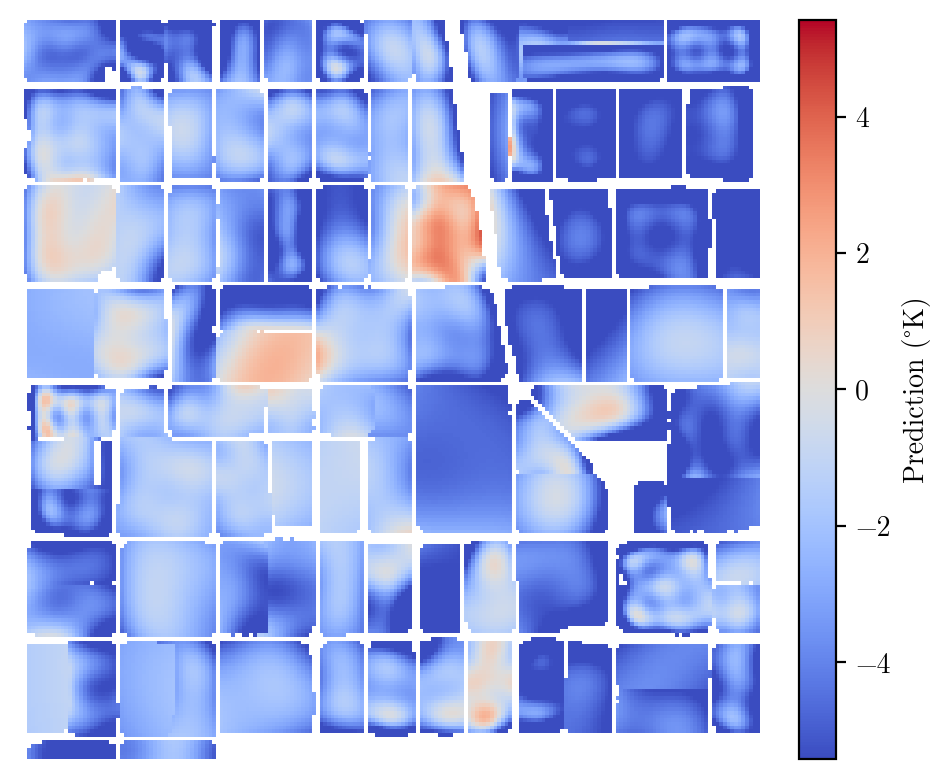}
    \caption{Kriging mean prediction}
    \label{fig:kriging_mean}
  \end{subfigure}
  \begin{subfigure}[b]{0.32\textwidth}
    \centering
    \includegraphics[width=\textwidth,height=4.5cm,keepaspectratio]{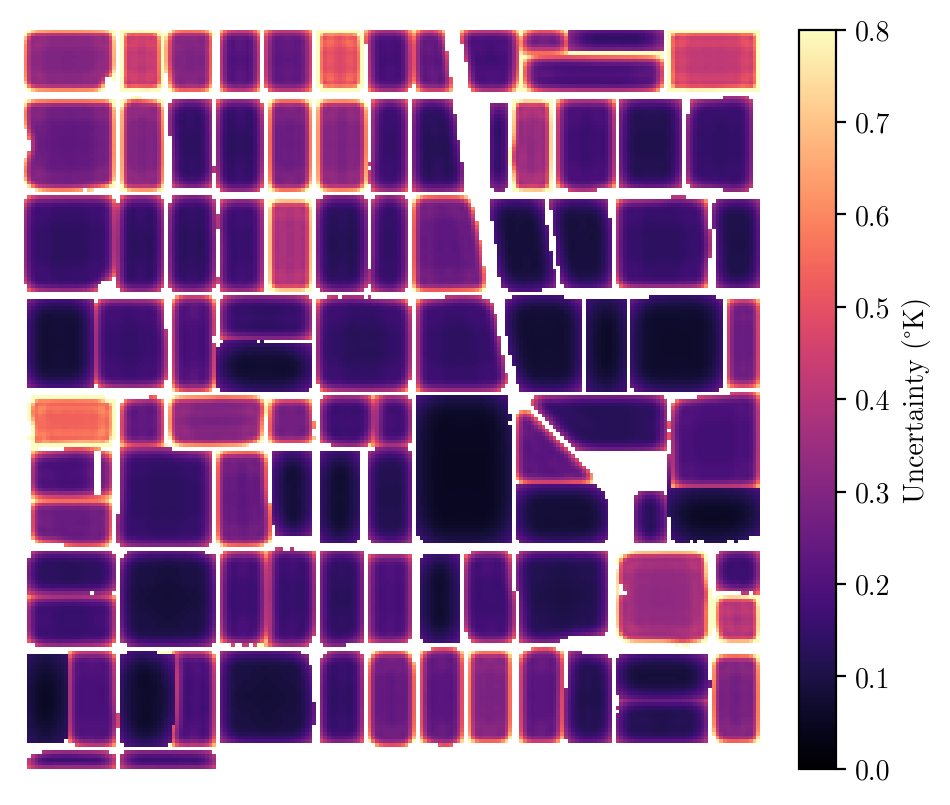}
    \caption{Kriging per-pixel $2 \sigma$}
    \label{fig:kriging_variance}
  \end{subfigure}
  \caption{Comparison of (a) the original ECOSTRESS residual image, (b) the kriging mean prediction, and (c) two per-pixel standard deviations for the kriging predictions.}
  \label{fig:deblurring_results}
\end{figure}

Next, we show the fitted parameters for each region in \Cref{fig:learned_parameters}. To estimate the parameters, we use 4 ECOSTRESS images (not including the target image) from 2020 to 2021, spaced by approximately 3 months. Similarly, we use 4 Landsat images starting from November 2021, separated by approximately 3 months each. We see that the length scales vary significantly across regions, justifying our use of a block-diagonal GP to model the agricultural field structure. Furthermore, the optimization procedure appears to have found reasonable parameter values for both datasets. Finally, we depict the kriging results along with the resulting uncertainty quantification in \Cref{fig:deblurring_results}. The kriging predictions effectively deblur many of the features present in the original ECOSTRESS image, and we can quantify the uncertainty in our predictions through the per-pixel variance. Note that the uncertainty for each region is higher near the edges; this is because the statistical assumption of independence between regions increases uncertainty of the reconstruction near the edges.

\vspace{.3cm}

\vspace{.3cm}

We verify that our method works well by manually blurring and recovering a high-resolution Landsat image in \Cref{app:verification}.

\section{Discussion} \label{sec:discussion}

In this work, we proposed a novel statistical method for downscaling LST data from the ECOSTRESS mission using high-resolution data from the Landsat mission as a proxy for modeling agricultural field structure. By leveraging a block-diagonal Gaussian process model, we were able to capture the spatial structure of agricultural fields and perform scalable inference. Our results demonstrate that the proposed method effectively deblurs ECOSTRESS LST data, producing higher-resolution estimates along with uncertainty quantification. The fitted parameters varied significantly across regions, justifying our use of a block-diagonal GP to model the agricultural field structure. Furthermore, our toy example verified that the method can successfully reconstruct blurred images, recovering features present in the original data.

Despite the strengths of our proposed framework, several limitations remain. First, while we estimate parameters for the network of roads between agricultural fields, we do not reconstruct this region in the current work due to its size. Second, despite the strong performance of the segmentation method using the Segment Anything Model (SAM), future studies might benefit from incorporating alternative segmentation algorithms or incorporating the uncertainty of the segmentation. Furthermore, the performance of our method is inherently limited by the quality of the ECOSTRESS data, which suffers from blur and artifacts resulting from physical damage to the sensor. This extends to our assumptions that the blur kernel and additive sensor noise are Gaussian; these assumptions may be inexact given that \citet{holmes2024orbit} demonstrate that the ECOSTRESS point spread function is not perfectly Gaussian. Finally, to ensure identifiability, we rely on the simplifying assumption that the length scale parameters for ECOSTRESS and Landsat are identical, a condition that may not strictly hold in practice.

In future work, the Block-Diagonal Gaussian Process (BDGP) kernel could be augmented with a temporal length scale to capture time-varying dynamics. This would allow for gap-filling in time, which is of great practical given the large amount of missingness. Additionally, replacing the standard squared exponential kernel with a sharper Mat\'ern kernel could more accurately model the roughness of the field. From a computational perspective, further optimizing the code would allow scalability to larger datasets. Another future direction is to model the mean of the high-resolution land surface temperature field, possibly using an auxiliary (higher-resolutiondata source. Finally, the roads could be partitioned into smaller, more tractable regions so that they may be reconstructed as well.

The broader implication of this work is that by modeling a field structure using a block-diagonal GP, we obtain a pipeline for building a scalable statistical model for data fusion of sensors with complementary strengths. This approach accounts for both the change of support and field structure present in the data, allowing us to perform scalable statistical modeling on large datasets. For instance, this method could be applied to urban heat island modeling, European agricultural landscapes (with more irregular field shapes), or other remote sensing contexts where high-resolution proxies are available.

\section*{Data availability}

Code can be found at \url{https://github.com/sanjitdp/downscaling-lst}. Replication data can be obtained from \url{https://appeears.earthdatacloud.nasa.gov/} using the \verb+.json+ request files contained in the code repository.

\printbibliography

\begin{appendices}
    \crefalias{section}{appendix}
    \section{Estimating the ECOSTRESS variance parameter} \label{app:variance-est}

  In this section, we justify \eqref{eq:variance-estimator} using elementary Fourier analysis (e.g., \cite{stoica2005spectral}). Assuming that there is no blurring at the edges, we can compute the power spectral density (PSD) of $f \coloneq f_{R_r}^\mathrm{ES}$ as
  \begin{align*}
    S_f(\omega) = A \exp(-4\pi^2 (\ell_r^\mathrm{ES})^2\, \norm{\omega}^2)
  \end{align*}
  for some constant $A > 0$. Similarly, we can compute the energy spectral density of the blur kernel $\kappa \coloneq \kappa_\mathrm{blur}$ as
  \begin{align*}
    \abs{\hat{\kappa}(\omega)}^2 = \exp(-4\pi^2 \sigma_\mathrm{blur}^2 \norm{\omega}^2),
  \end{align*}
  where $\hat{\kappa}$ denotes the Fourier transform of $\kappa$. The PSD of the convolution factorizes as a product, so we obtain
  \begin{align*}
    S_{f * \kappa}(\omega)
    = S_f(\omega)\, \abs{\hat{\kappa}(\omega)}^2
    = A \exp\left( -4\pi^2 \left( (\ell_r^\mathrm{ES})^2 + \sigma_\mathrm{blur}^2 \right) \norm{\omega}^2 \right).
  \end{align*}
  In particular, integrating the PSD over all frequencies yields the variance:
  \begin{align*}
    \Var(f \ast \kappa)
     & = \iint S_{f * \kappa}(\omega)\, d\omega                                                                                      \\
     & = \iint A \exp\left( -4\pi^2 \left( (\ell_r^\mathrm{ES})^2 + \sigma_\mathrm{blur}^2 \right) \norm{\omega}^2 \right)\, d\omega \\
     & = \frac{A}{4\pi \left( (\ell_r^\mathrm{ES})^2 + \sigma_\mathrm{blur}^2 \right)}.
  \end{align*}
  Similarly, we have
  \begin{align*}
    (\sigma_r^\mathrm{ES})^2
    \coloneq \Var(f)
    = \iint S_f(\omega)\, d\omega
    = \iint A \exp(-4\pi^2 (\ell_r^\mathrm{ES})^2\, \norm{\omega}^2)\, d\omega
    = \frac{A}{4\pi (\ell_r^\mathrm{ES})^2}.
  \end{align*}
  Note that $\Var(y_{R_r}^\mathrm{ES}) = \Var(f \ast \kappa) + \sigma_\mathrm{sensor}^2$ due to independence between the (blurry) signal and the sensor noise. Putting the pieces together, we obtain
  \begin{align*}
    (\sigma_r^\mathrm{ES})^2
     & = \Var(f * \kappa) \left( \frac{(\ell_r^\mathrm{ES})^2 + \sigma_\mathrm{blur}^2}{(\ell_r^\mathrm{ES})^2} \right)                                    \\
     & = (\Var(y_{R_r}^\mathrm{ES}) - \sigma_\mathrm{sensor}^2) \left( \frac{(\ell_r^\mathrm{ES})^2 + \sigma_\mathrm{blur}^2}{(\ell_r^\mathrm{ES})^2} \right).
  \end{align*}
  Finally, the empirical variance $\widehat{\Var}(y_{R_r}^\mathrm{ES})$ is a consistent estimator of $\Var(y_{R_r}^\mathrm{ES})$ because the kernel is squared exponential (and hence stationary and ergodic). Thus, it follows that the proposed estimator \eqref{eq:variance-estimator} for $\sigma_r^\mathrm{ES}$ is consistent.



\section{Additional figures} \label{app:additional-figures}

\begin{figure}[htbp]
  \centering
  \begin{subfigure}[b]{0.48\textwidth}
    \centering
    \includegraphics[width=\textwidth,height=4.5cm,keepaspectratio]{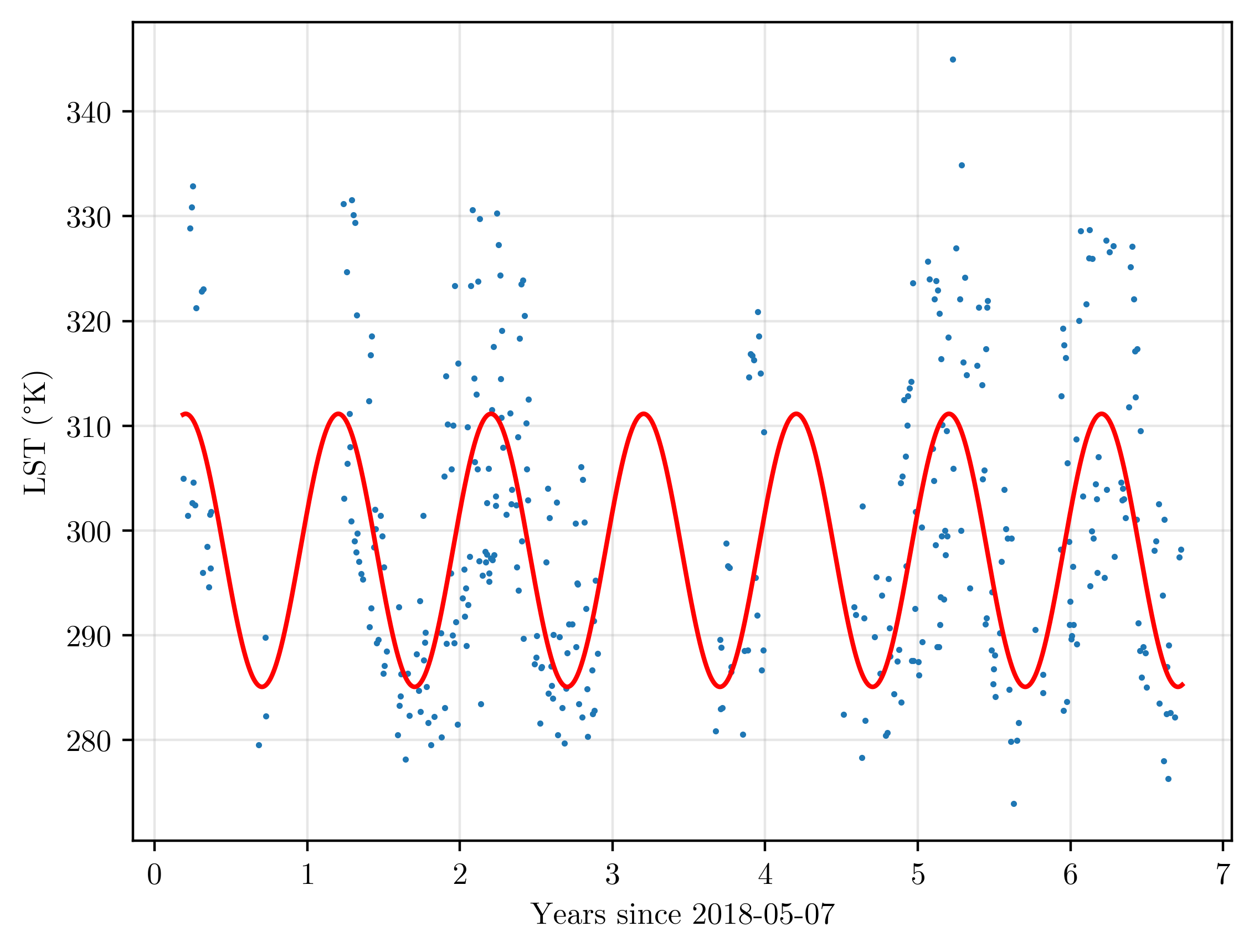}
    \caption{Annual cycle}
    \label{fig:annual_cycle}
  \end{subfigure}
  \begin{subfigure}[b]{0.48\textwidth}
    \centering
    \includegraphics[width=\textwidth,height=4.5cm,keepaspectratio]{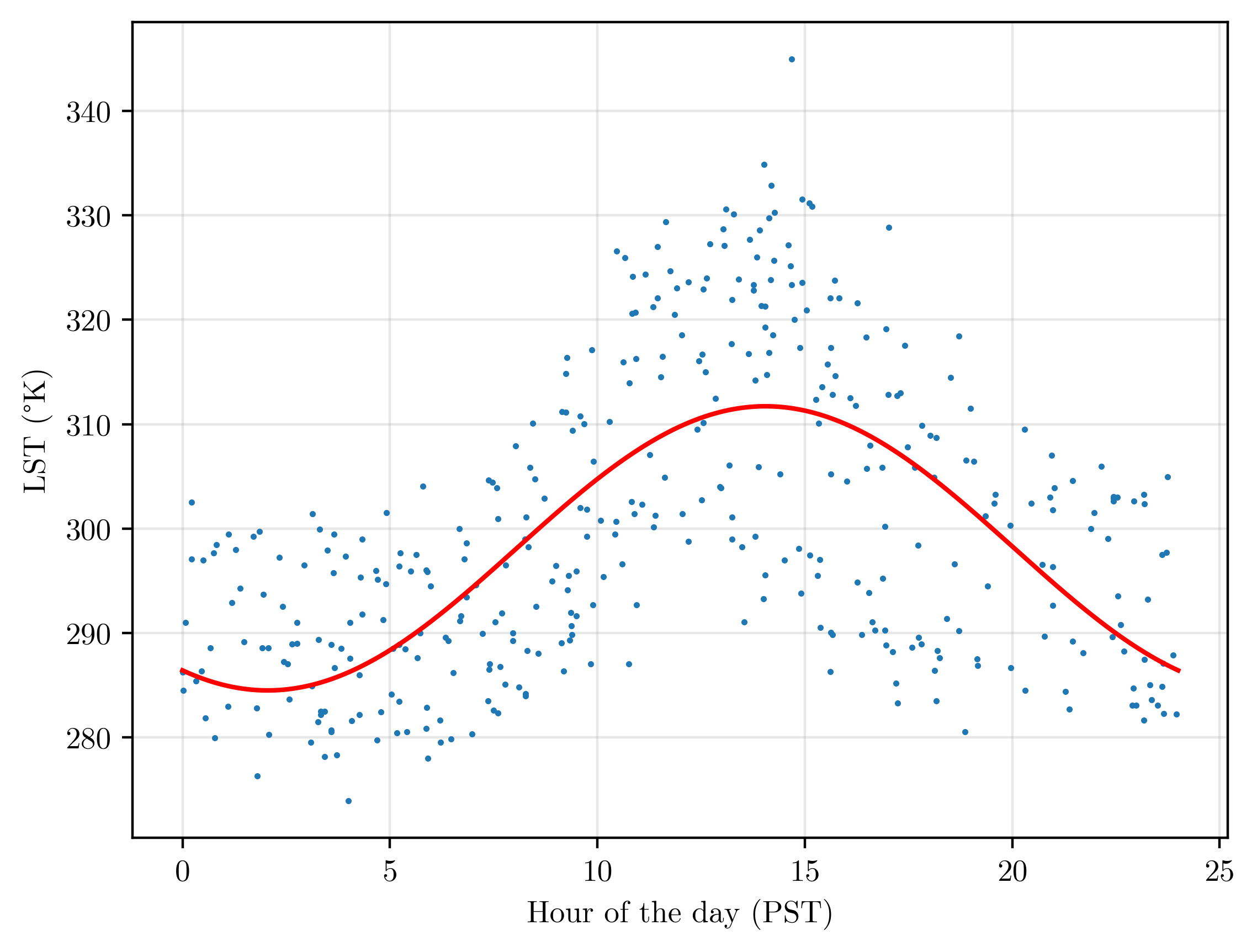}
    \caption{Diurnal cycle}
    \label{fig:diurnal_cycle}
  \end{subfigure}
  \caption{Plots of the fitted temperature cycle at a randomly selected pixel for (a) the annual cycle from ECOSTRESS data and (b) the diurnal cycle from ECOSTRESS data.}
  \label{fig:fourier_fit_pixel}
\end{figure}

\begin{figure}[htbp]
  \centering
  \includegraphics[width=0.96\textwidth,height=4.5cm,keepaspectratio]{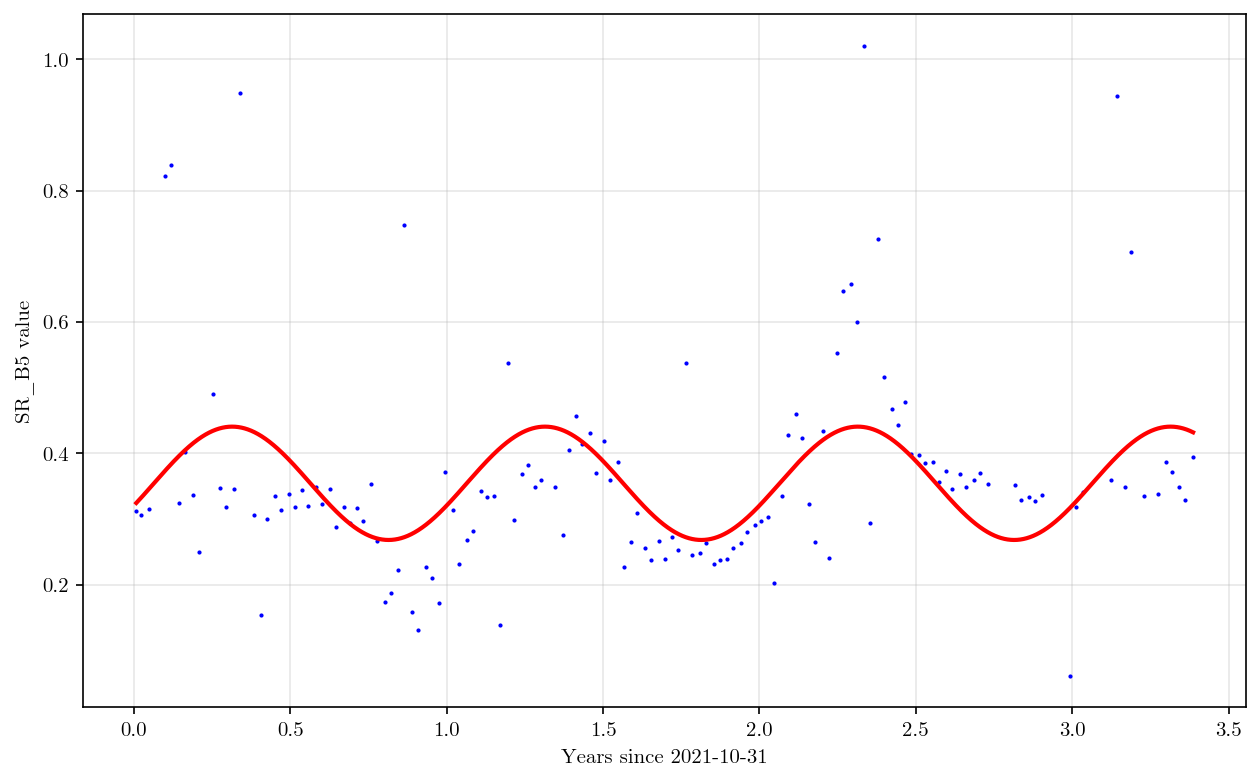}
  \caption{Plot of the annual cycle of the SR\_B5 channel of the Landsat data at a randomly selected pixel. This channel measures the surface reflectance in the near-infrared spectrum of light.}
  \label{fig:annual_cycle_landsat}
\end{figure}

In this appendix, we collect several additional figures (\Cref{fig:fourier_fit_pixel} and \Cref{fig:annual_cycle_landsat}).

  \section{Verification of methods} \label{app:verification}

\begin{figure}[htbp]
  \centering
  \begin{subfigure}[b]{0.48\textwidth}
    \centering
    \includegraphics[width=\textwidth,height=4.5cm,keepaspectratio]{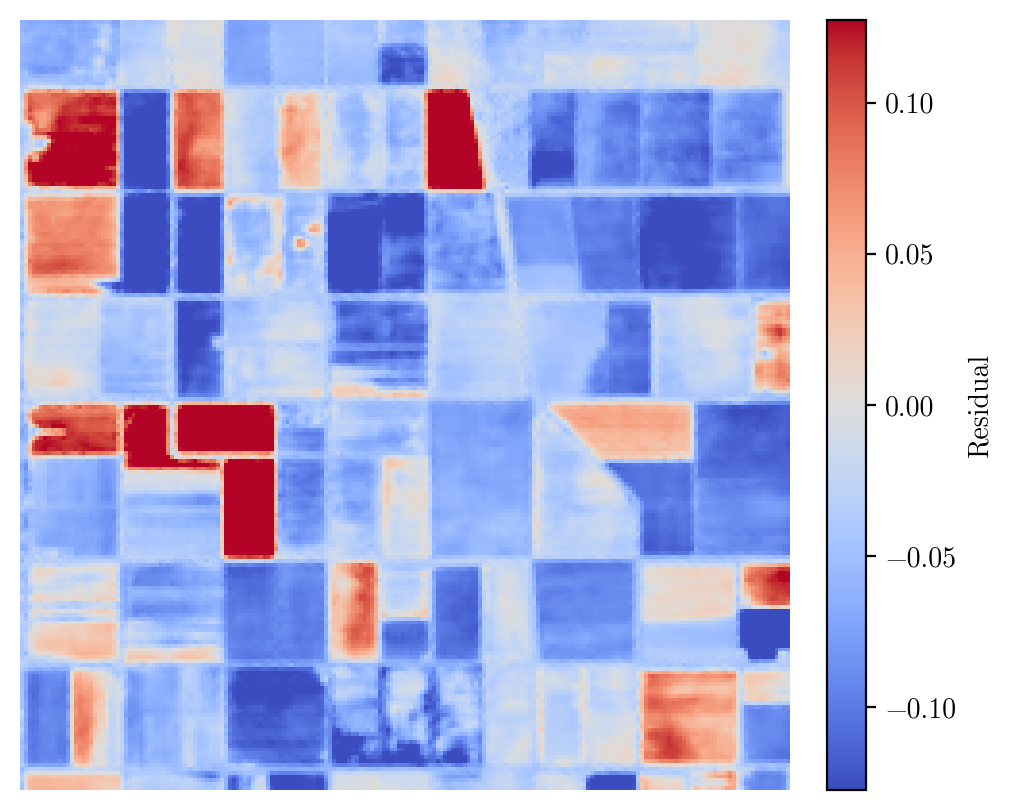}
    \caption{Original image}
    \label{fig:toy_original}
  \end{subfigure}
  \begin{subfigure}[b]{0.48\textwidth}
    \centering
    \includegraphics[width=\textwidth,height=4.5cm,keepaspectratio]{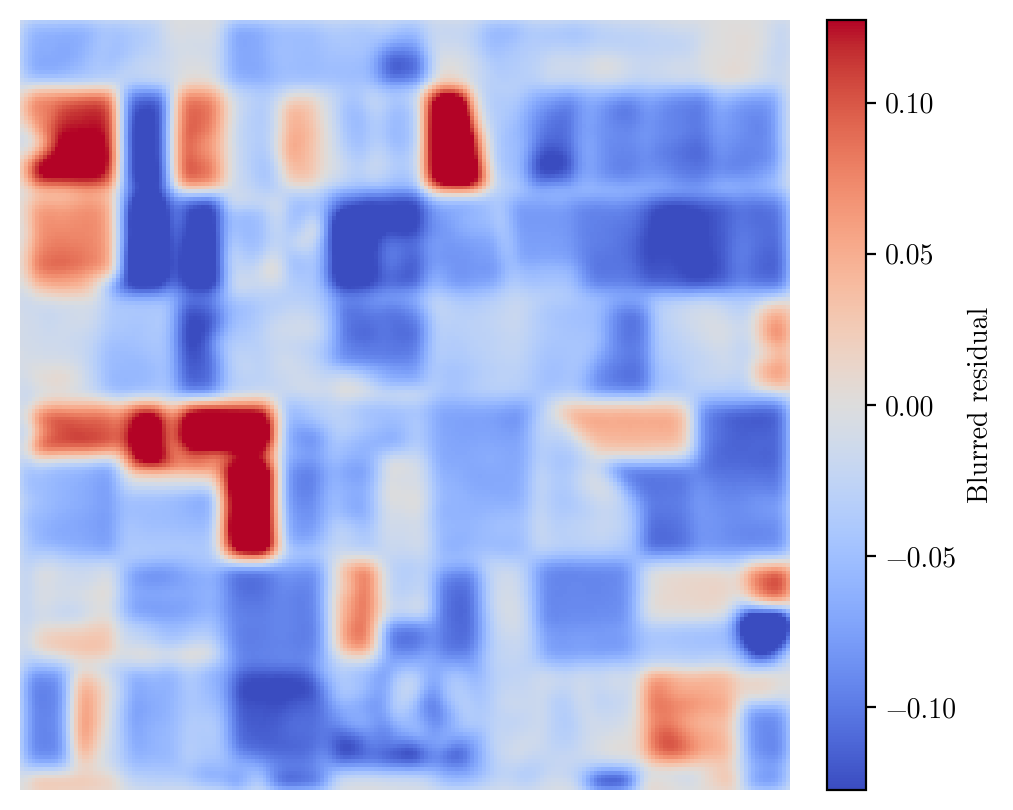}
    \caption{Blurred image}
    \label{fig:toy_blurred}
  \end{subfigure}
  \caption{(a) the original high resolution image and (b) the manually blurred image used as input to the toy example.}
  \label{fig:toy_input}
\end{figure}

\begin{figure}[htbp]
  \centering
  \begin{subfigure}[b]{0.48\textwidth}
    \centering
    \includegraphics[width=\textwidth,height=4.5cm,keepaspectratio]{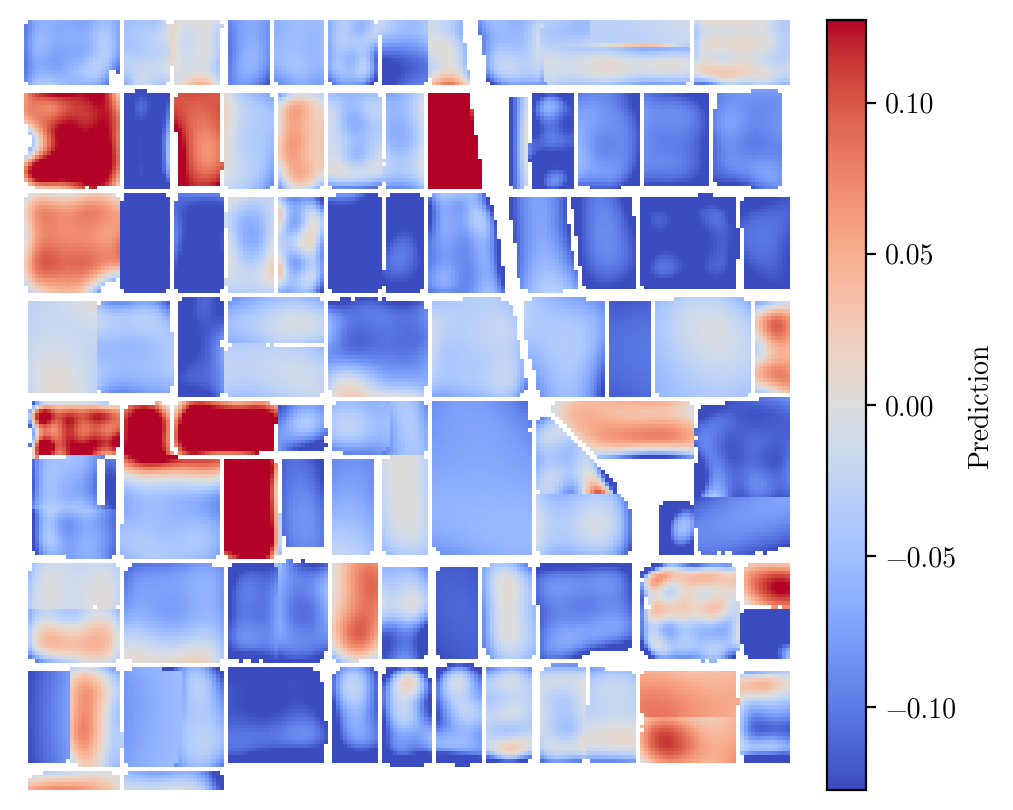}
    \caption{Kriging mean}
    \label{fig:toy_kriging_mean}
  \end{subfigure}
  \begin{subfigure}[b]{0.48\textwidth}
    \centering
    \includegraphics[width=\textwidth,height=4.5cm,keepaspectratio]{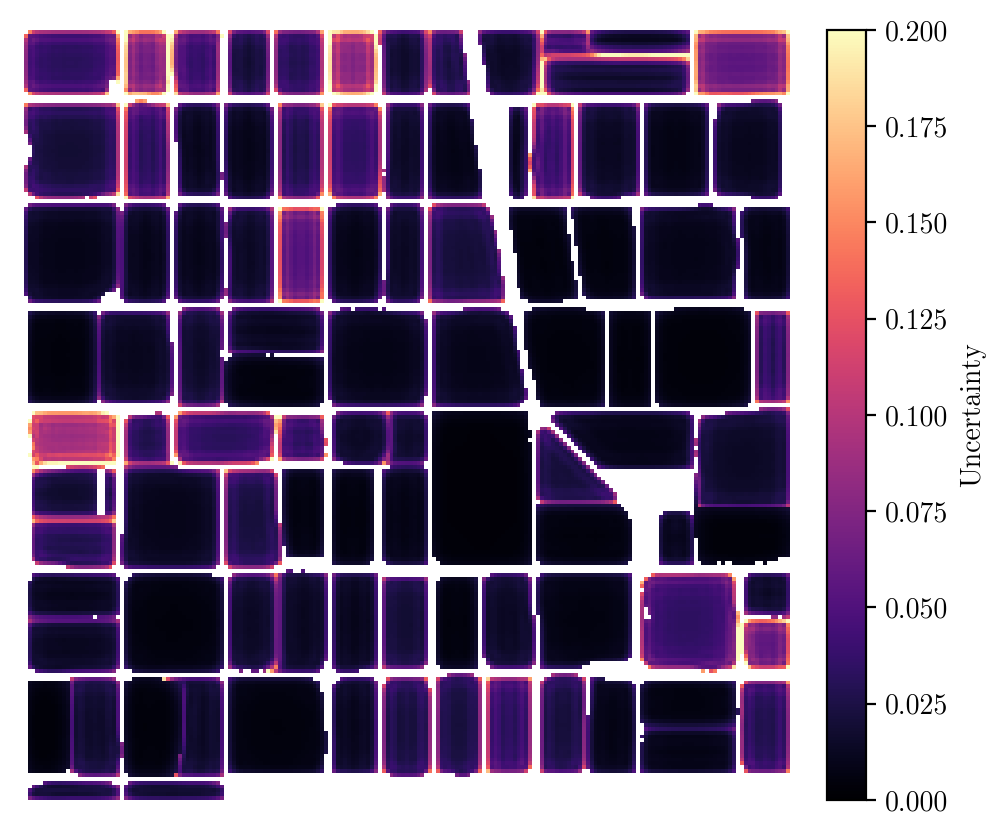}
    \caption{Kriging per-pixel $2 \sigma$}
    \label{fig:toy_uncertainty}
  \end{subfigure}
  \caption{Verification of the proposed method on the toy example of \Cref{fig:toy_input}: (a) the kriging mean prediction and (b) two per-pixel standard deviations for the kriging predictions.}
  \label{fig:toy_output}
\end{figure}

  To verify that our method works, we conclude by manually blurring a high-resolution Landsat image (\Cref{fig:toy_input}). We set $\sigma_\mathrm{blur}$ to 0.97 ECOSTRESS pixels (70 m), which is the same as the amount of blurring that we found to be present in the ECOSTRESS dataset. Then, we set $\sigma_\mathrm{sensor} = 10^{-4}$, which is reasonable because we expect much smaller sensor noise in Landsat compared to ECOSTRESS. We use the same estimated parameters from the real data in this toy example. Using these parameters, we reconstruct the image using a simplified version of our pipeline (\Cref{fig:toy_output}). Note that the kriging mean does de-blur the image, recovering some of the features present in the original image. Our choice of a squared exponential kernel results in a relatively smooth reconstruction; one could obtain a rougher reconstruction using a different kernel, but we leave this for future work.

\section{Refining the oversegmented image} \label{app:combination-algorithm}

In this appendix, we provide our algorithm to refine the oversegmented image provided by the Segment Anything Model (SAM).

\begin{itemize}
  \item Discard all segments with fewer than 100 pixels.
  \item Sort the remaining segments in increasing order of area.
  \item Iterate through the segments $s$ from largest to smallest:
        \begin{itemize}
          \item Subtract $s$ from all previously kept segments wherever they overlap.
          \item Add the current segment to the list of kept segments.
        \end{itemize}
\end{itemize}
\end{appendices}

\end{document}